\documentclass[usegraphicx]{mn2e}
\usepackage{subfig}
\usepackage{epsfig}
\usepackage{amsmath}
\usepackage{amssymb}
\usepackage{graphicx}
\usepackage{times,longtable,color}

\usepackage{wrapfig}

\newcommand {\apgt} {\ {\raise-.5ex\hbox{$\buildrel>\over\sim$}}\ }
\newcommand {\aplt} {\ {\raise-.5ex\hbox{$\buildrel<\over\sim$}}\ }

\title[Geometrical beaming of stellar mass ULXs]
{Geometrical beaming of stellar mass ULXs}

\author[M. Middleton \& A. King]
{Matthew J. Middleton$^{1}$ and Andrew King$^{2,3,4}$\\
\\
1. Institute of Astronomy, Madingley Road, Cambridge CB3 OHA\\
2. Theoretical Astrophysics Group, University of Leicester, Leicester LE1 7RH\\
3. Anton Pannekoek Institute, University of Amsterdam, Science Park 904, 1098 XH Amsterdam, Netherlands\\ 
4. Leiden Observatory, Leiden University, Niels Bohrweg 2, NL-2333 CA Leiden, Netherlands\\  
}

\pagerange{\pageref{firstpage}--\pageref{lastpage}} \pubyear{2011}
\long\def\symbolfootnote[#1]#2{\begingroup\def\thefootnote{\fnsymbol{footnote}}\footnote[#1]{#2}\endgroup} 
\def\ga{\mathrel{\hbox{\rlap{\hbox{\lower4pt\hbox{$\sim$}}}{\raise2pt\hbox{$>$}}
}}}

\begin{document}

\topmargin = -0.5cm

\maketitle

\label{firstpage}

\begin{abstract}
The presence or lack of eclipses in the X-ray lightcurves of ultraluminous X-ray sources (ULXs) can be directly linked to the accreting system geometry.  In the case where the compact object is stellar mass and radiates isotropically, we should expect eclipses by a main-sequence to sub-giant secondary star on the recurrence timescale of hours to days. X-ray lightcurves are now available for large numbers of ULXs as a result of the latest {\it XMM-Newton} catalogue. We determine the amount of fractional variability that should be injected into an otherwise featureless lightcurve for a given set of system parameters as a result of eclipses and compare this to the available data. We find that the vast majority of sources for which the variability has been measured to be non-zero and for which available observations meet the criteria for eclipse searches, have fractional variabilities which are too low to derive from eclipses and so must be viewed such that $\theta \le$ cos$^{-1}(R_{*}/a$). This would require that the disc subtends a larger angle than that of the secondary star and is therefore consistent with a conical outflow formed from super-critical accretion rates and implies some level of geometrical beaming in ULXs.
\end{abstract}

\section{Introduction}

There is mounting evidence that the global population of ULXs is heterogenous, with the discovery of sources that show both spectral and temporal behaviour that does not obviously conform to that of the wider population (see Gladstone et al. 2009; Sutton, Roberts \& Middleton 2013; Middleton et al. 2015a). Notably amongst the outliers are M82 X-2 which shows pulsations in {\it NuSTAR} data and identifies the compact object as a neutron star (Bachetti et al. 2014), and HLX-1, the spectral evolution of which appears analogous to black hole binaries in outburst (e.g. Fender, Belloni \& Gallo 2004) and has to-date provided the strongest evidence for an IMBH (M $>$ 1000s M$_{\odot}$) outside of dwarf AGN (Farrell et al. 2009; Davis et al. 2011; Servillat et al. 2011 but see also King \& Lasota 2014; Lasota et al. 2015). The remainder of the population are likely to contain stellar remnants with masses $<$100 M$_{\odot}$ and, given their remarkable X-ray luminosities of $>$ 1$\times$10$^{39}$ erg s$^{-1}$, provide strong evidence for the presence of super-critical accretion likely due to high mass transfer rates via Roche lobe overflow (RLO) from a more massive companion star in a tight binary orbit (see for example the case of SS433: King et al. 2000; Begelman, King \& Pringle, 2006). 

The picture of ULXs as super-critical accretors has received a recent boost with a dynamical mass determination for one source (Motch et al. 2014), and the discovery of winds from two archetypal ULXs (Pinto, Middleton \& Fabian 2016, see also Middleton et al. 2014; 2015b) with outflow velocities of $\ge$ 0.2c. The absorption features that indicate the presence of a mass-loaded outflow require that we view into the wind at least some of the time (see Middleton et al. 2015b) but the opening angle and homogeneity of the wind is still unknown and is important for constraining the impact of geometrical beaming (King 2009) and the origin of variability (Heil et al. 2009; Middleton et al. 2011; 2015b). However, it is important to note that the apparent wind properties agree well with expectations for strongly supercritical accretion (King \& Muldrew, 2016). 

Eclipses have been previously identified as a useful diagnostic of ULXs by Pooley \& Rappaport (2005) and more recently by King \& Nixon (2016); where no eclipses are found on the recurrence timescales of hours to days, this would imply that a main-sequence to sub-giant secondary star does not transit the stellar mass source, whilst on recurrence timescales of months to years, the case for an IMBH being eclipsed can be tested. The latter is still extremely difficult to demonstrate unambiguously due to the high cadence of observations required, however, the availability of X-ray lightcurves spanning hours to days is now available for many ULXs (Rosen et al. 2015) allowing us to test the former, stellar-mass compact object scenario. Here we place limits on the likely presence of eclipses in a large sample of ULXs using a large test range in binary separation and mass ratio, the mass limit placed by Motch et al. of $<$ 15 M$_{\odot}$ and the impact that eclipses must have on the variability. 

\begin{figure*} 
 \centering
\includegraphics[width=12cm]{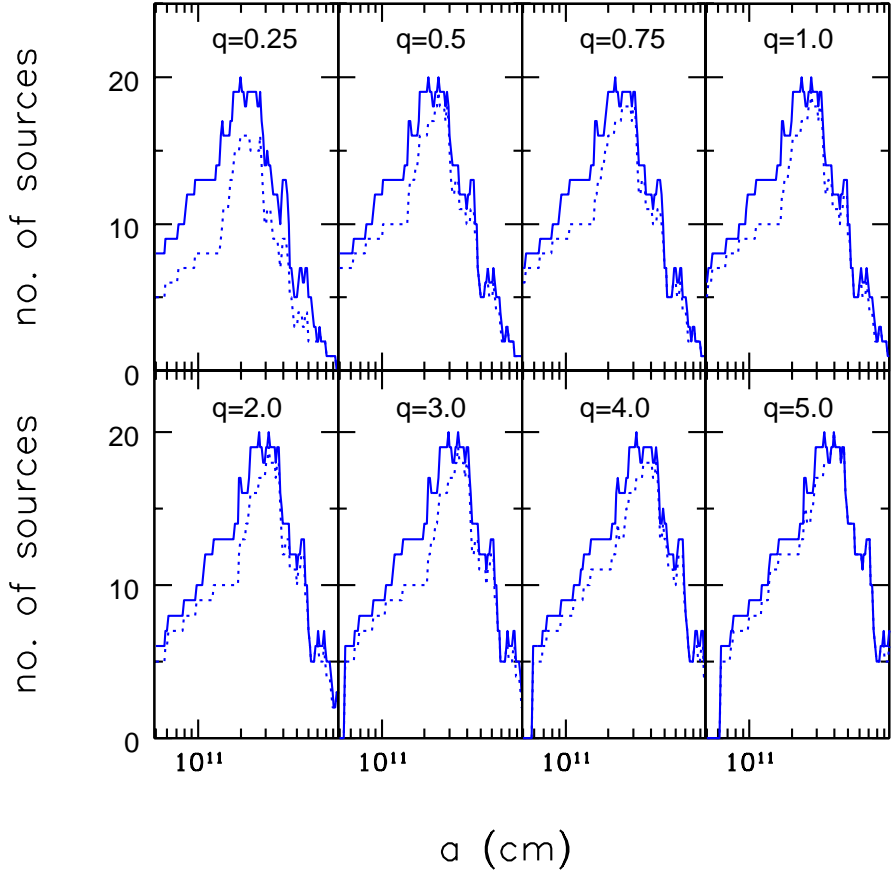} 
            \caption{The total number of individual sources (solid line) where an observation satisfies the selection criteria in the text as a function of the binary separation ($a$) and mass ratio ($q$). The dotted line indicates how many of these sources also satisfy the criteria in equation 7 and are therefore consistent with not showing eclipses.}
  \label{fig:example}%
\end{figure*}

\section{Constraining the presence of eclipses}

Whilst eclipses have not yet been conclusively detected in ULXs (with one recent exception: Urquhart \& Soria 2016 submitted) - although dip like events have been seen in a number of sources (Stobbart, Roberts \& Warwick 2004; Grise et al. 2013) - we can consider the impact that eclipses by an opaque body must have on the lightcurve of a given source for a range of system parameters. The summed variance resulting {\it only} from an eclipse of duration T$_{\rm e}$ occurring in an observation of length T$_{\rm o}$ by a passing opaque star of radius R$_{*}$ projected as a perfect disc is:

\begin{equation}
\sigma^2 = \frac{NT_{e}}{T_{o}}\mu^{2} + \left(1-\frac{NT_{e}}{T_{o}}\right)(x-\mu)^{2}
\end{equation}

\noindent or, in terms of fractional variability:

\begin{equation}
\left(\frac{\sigma}{\mu}\right)^{2} = \frac{NT_{e}}{T_{o}} + \left(1-\frac{NT_{e}}{T_{o}}\right)(\frac{x}{\mu}-1)^{2}
\end{equation}

\noindent where $\mu$ is the mean count rate, x is the count rate out of eclipse (in eclipse the count rate is assumed to be zero, i.e. the eclipse is of a point source) and N is the number of eclipses occurring in a given observation (i.e. T/T$_{\rm o}$ where T is the binary period). We have also assumed that the X-ray source is compact enough such that the duration of ingress and egress can be ignored. This formula can be simplified and re-arranged to:

\begin{equation}
T_{e} = \frac{T_{o}}{N}\left(1-\left[\left(\frac{\sigma}{\mu}\right)^{2} + 1\right]^{-1}\right)
\end{equation}

\noindent These formulae make the implicit assumption that there are no timebins which straddle the ingress and egress -  we will return to this assumption when discussing the real data (\S 3.1).

The eclipse duration is also a well understood function of the stellar radius ($R_{*}$), binary separation ($a$), $T$ and the inclination of the system ($\theta$, which is taken to be from the axis perpendicular to the binary system which we assume to be perfectly aligned: see King \& Nixon 2016 for a discussion on probable inclination evolution in ULXs):

\begin{equation}
T_{e} = \frac{1}{\pi}\left(\frac{R_{*}}{a}\right)T\rm{sin}(\theta)
\end{equation}

Assuming stable RLO we can substitute for $TR_{*}/a$ (Frank, King \& Raine 2002) using:

\begin {equation}
\frac{R_{*}}{a} = \frac{0.49 q^{2/3}}{0.6 q^{2/3} + \rm{ln}\left(1+q^{1/3}\right)}
\end{equation}

\noindent and

\begin {equation}
T = \left[\frac{a}{3\times10^{11} M_{1}^{1/3}\left(1+q\right)^{1/3}}\right]^{3/2} {\rm days}
\end{equation}

Using equation 3 we can determine the implied eclipse duration based on a {\it measured} fractional variability; using this along with values for $R_{*}/a$ and $T$ determined from values for $q$ and $a$ we can then determine $\theta$ from a re-arranged equation 4. If this value is smaller than cos$^{-1}(R_{*}/a)$ then, by definition, no eclipses are possible.

\section{Utilising the catalogues}

Walton et al. (2011) combined the {\it XMM-Newton} 2XMM source catalogue with the RC3 (de Vaucouleurs et al. 1991) and Karachentsev et al. (2004) nearby galaxy catalogues to identify a total of 470 ULX candidates in the local Universe (out to a distance of 148 Mpc) with a sky coverage of 504 square degrees, correcting for background contaminants. The latest incarnation of the {\it XMM-Newton} source catalogue (3XMM-DR5: Rosen et al. 2015) provides the excess fractional variability (i.e. the Poisson noise is subtracted) from the 0.2-12~keV lightcurve determined using a bin-size ($\Delta t$) corresponding to 18 ct/s across the entire duration of the observation (minus intervals not defined by a GTI and null bins). 

By cross-matching 3XMM-DR5 with the Walton et al. (2011) ULX catalogue we have obtained all existing observations of the previously identified ULX candidates using {\it XMM-Newton} which includes the observation length ($T_{\rm o}$) and fractional variability ($\sigma/\mu$). We restrict ourselves to use of the PN only, as typically the fractional rms will be better constrained as the Poisson noise is lower due to higher detector throughput (see Vaughan et al. 2003); we therefore exclude observations where the PN was not available and where the fractional variability was not determined - this left us with a total of 123 individual sources across 223 observations. 

Assuming the mass of the compact object to be $<$15M$_{\odot}$ and therefore consistent with the findings of Motch et al. (2014), we can determine the fraction of sources where eclipses should not be possible, i.e. they meet the criterion:

\begin {equation}
\rm{sin}^{-1}\left(\frac{\pi a T_{e}}{R_{*}T}\right) \le \rm{cos}^{-1}\left(\frac{R_{*}}{a}\right) 
\end {equation}

\noindent We do this across a range in binary separation of 5$\times$10$^{10}$ -  1$\times$10$^{12}$ cm and for mass ratios of 0.25-5. We place sensible constraints of $T \ge$ 0.01 days, $\Delta t \le T_{\rm e}/2$ and T$_{\rm o}$ $>$ T; the first restriction comes from rough observational limits for binaries containing neutron stars and black holes, and the second implies that there is always a full bin in the eclipse with zero count/s. From equation 6 it is clear that the number of observations fulfilling the third criterion varies as a function of binary separation and mass ratio but is typically Gaussian ($\gtrsim$ 20 observations) except for large $a$. In Figure 1 we plot the number of individual sources (rather than observations which is naturally larger) which meet our selection criteria and the number which also meet the criterion in equation 7 as a function of $q$ and $a$.



\subsection{Caveat on $\sigma/\mu$}

A notable caveat in this work is the assumption that the fractional variability (equation 3) is not affected by bins which straddle the ingress and egress. This is highly idealised and in reality such occurrences will lower the effective rms introduced by the eclipse. We can however determine the likely impact such straddling can have by varying the sampling of the eclipse, i.e. by varying the relative fraction of the straddling bin which falls inside or outside of the eclipse we can determine, based on the binsize, eclipse duration and number of eclipses in a given observation length, the variance relative to the idealised case in equation 1. We do this for all observations that meet the criteria discussed above and plot the mean fractional difference in variance as a function of $q$ and $a$ in Figure 2. In order to simplify the error determination, we assume only integer bins across the observation which is not what we assume in the previous section, thus the derived error is only approximate. As the mean count rate is unchanged by the sampling, the fractional error in the variance is a measure of the fractional error on $(\sigma/\mu)^{2}$ in equation 3. The result of our calculation indicates that at most we should expect an approximate error on the variance of $<$ 20\%, converting this to an error on the eclipse duration we find that $dT_{\rm e}/T_{\rm e} = -d\sigma^{2}/[\sigma^{2}(\sigma^{2}+1)]$ so that the fractional error in the eclipse duration is always smaller than that in the variance estimate. Thus we do not expect a significant impact on the derived inclination of the source and the overall result.

\begin{figure} 
 \centering
\includegraphics[width=8cm]{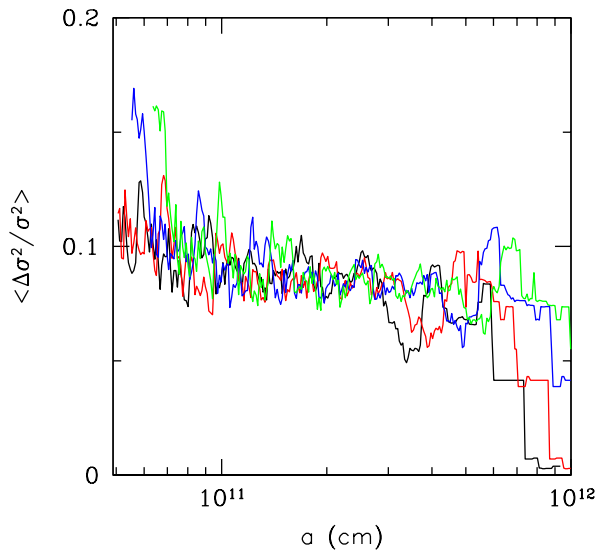} 
            \caption{Estimate of the mean fractional error in the variance as a result of including timebins which fall over the ingress and egress of an otherwise flat lightcurve with square eclipses. The colour scheme corresponds to mass ratios of $q$=0.25 (black), $q$=1 (red), $q$=3 (blue) and $q$=5 (green).}%
\end{figure}


\section{Discussion \& Conclusion}

If we assume that the mass determined by Motch et al. (2014) of $<$ 15 M$_{\odot}$ is representative for those sources which show `ultraluminous' spectra (Stobbart et al. 2006; Gladstone et al. 2009; Sutton et al. 2013), then assuming material is transferred via RLO, the variability for the vast majority of the best available sample (following our selection criteria) is consistent with the lack of eclipses by a companion star (in a fully aligned orbit). In the picture where a super-critical accretion rate leads to a large radiation pressure and an inflated disc (e.g. Shakura \& Sunyaev 1973; King 2001, 2009; Poutanen 2007; Ohsuga \& Mineshige 2011; Jiang et al. 2014), then the secondary star is expected to be fully obscured by the large scale-height inflow and radiatively driven wind (H/R $>$ 1). This picture leads to a conical geometry where, in order to see the X-ray emission at all and define the source as a ULX, we must generally be viewing into the cone or through the wind where down-scattering does not entirely prevent some bright X-ray emission. The lack of large contributions to the variability via eclipses is therefore fully consistent with this picture and supports arguments built on spectral-variability properties (Sutton et al. 2013; Middleton et al. 2015a), the detection of blue-shifted absorption features (Middleton et al. 2014; 2015b; Pinto et al. 2016) and implies that some level of geometrical beaming is likely (probably only a factor of a few at most given the ionising luminosity incident upon nebulae surrounding some of the sources: Pakull \& Mirioni 2003). 

In this work we have assumed that the rms values are representative (i.e. we have ignored the errors on the rms values); whilst uncertainty on the rms can introduce a corresponding uncertainty on the derived inclination this is somewhat mitigated by the assumption that {\it all} the variability measured in a given observation comes from an eclipse which we know to be incorrect due to the observation of band-limited power spectra and rms-flux relationships (e.g. Heil et al. 2009; 2010) which implies that propagation of surface density fluctuations (Lyubarskii 1997; Arevalo \& Uttley 2006; Ingram \& Done 2011) is responsible for much of the variability (see Middleton et al. 2015a for possible means to extract variability from super-critical flows). As a result, the true variability imprinted by an eclipse could be considerably smaller than implied by the rms value we have used.


Our sample has not been selected based on spectral type nor luminosity (except that the inferred {\it isotropic} luminosity has at some point reached or exceeded 1$\times$10$^{39}$ erg s$^{-1}$ such that it has been identified as a ULX candidate: Walton et al. 2011). A lack of explicit selection might present an issue as there is a clear dichotomy in the general population, with the fainter ULXs likely associated with more `standard' accretion in a HM or LMXB proceeding around the Eddington limit (e.g. Middleton et al. 2011; 2012; 2013; Soria et al. 2012) whilst the brighter sources are more likely associated with thermal timescale mass transfer proceeding at super-critical rates (King 2001). However, we {\it have} selected on variability, which is known to be a tracer of  ULX `type' (see Sutton, Roberts  \& Middleton 2013; Middleton et al. 2015a) with only the latter, super-critical candidates (which are often the brighter sources but not always, see Middleton et al. 2015a) showing considerable fractional variability. 



Finally, an inescapable question that results from identifying ULXs as relatively low inclination sources is ``where are the edge-on ULXs?"; the answer is that the wind and the large-scale height disc will increasingly block the hard X-ray emission from the inner regions such that the sources become relatively X-ray faint e.g. NGC 55 ULX-1 (Middleton et al. 2015a) and may no longer appear above 1$\times$10$^{39}$ erg s$^{-1}$ or even X-ray bright at all. Instead, the photosphere we see to peak in the soft X-rays in ULXs may not be the photosphere we see at very high inclinations (Poutanen et al. 2007) and the X-ray emission incident onto the wind and disc will likely be reprocessed down to the UV (depending on the density along the line-of-sight for Compton scattering).


\section{Acknowledgements}

We thank the referee for their useful suggestions and Will Alston for helpful discussion. MJM appreciates support from an Ernest Rutherford STFC fellowship. This work is based on observations obtained
with {\it XMM-Newton}, an ESA science mission with instruments and
contributions directly funded by ESA Member States and NASA.

\label{lastpage}


\begin{thebibliography} {}

\bibitem[\protect\citeauthoryear{Ar{\'e}valo \& Uttley}{2006}]{2006MNRAS.367..801A} Ar{\'e}valo P., Uttley P., 2006, MNRAS, 367, 801
\bibitem[\protect\citeauthoryear{Bachetti et al.}{2014}]{2014Natur.514..202B} Bachetti M., et al., 2014, Natur, 514, 202
\bibitem[\protect\citeauthoryear{Begelman, King, \& Pringle}{2006}]{2006MNRAS.370..399B} Begelman M.~C., King A.~R., Pringle J.~E., 2006, MNRAS, 370, 399
\bibitem[\protect\citeauthoryear{Davis et al.}{2011}]{2011ApJ...734..111D} Davis S.~W., Narayan R., Zhu Y., Barret D., Farrell S.~A., Godet O., Servillat M., Webb N.~A., 2011, ApJ, 734, 111 
\bibitem[\protect\citeauthoryear{de Vaucouleurs et al.}{1991}]{1991rc3..book.....D} de Vaucouleurs G., de Vaucouleurs A., Corwin H.~G., Jr., Buta R.~J., Paturel G., Fouqu{\'e} P., 1991, rc3..book, I, 
\bibitem[\protect\citeauthoryear{D{\'{\i}}az Trigo et al.}{2006}]{2006A&A...445..179D} D{\'{\i}}az Trigo M., Parmar A.~N., Boirin L., M{\'e}ndez M., Kaastra J.~S., 2006, A\&A, 445, 179 
\bibitem[\protect\citeauthoryear{Farrell et al.}{2009}]{2009Natur.460...73F} Farrell S.~A., Webb N.~A., Barret D., Godet O., Rodrigues J.~M., 2009, Natur, 460, 73 
\bibitem[\protect\citeauthoryear{Fender, Belloni, \& Gallo}{2004}]{2004MNRAS.355.1105F} Fender R.~P., Belloni T.~M., Gallo E., 2004, MNRAS, 355, 1105
\bibitem[\protect\citeauthoryear{Frank, King, \& Raine}{2002}]{2002apa..book.....F} Frank J., King A., Raine D.~J., 2002, apa..book, 398
\bibitem[\protect\citeauthoryear{Gladstone, Roberts, \& Done}{2009}]{2009MNRAS.397.1836G} Gladstone J.~C., Roberts T.~P., Done C., 2009, MNRAS, 397, 1836
\bibitem[\protect\citeauthoryear{Gris{\'e} et al.}{2013}]{2013MNRAS.433.1023G} Gris{\'e} F., Kaaret P., Corbel S., Cseh D., Feng H., 2013, MNRAS, 433, 1023
\bibitem[\protect\citeauthoryear{Heil \& Vaughan}{2010}]{2010MNRAS.405L..86H} Heil L.~M., Vaughan S., 2010, MNRAS, 405, L86
\bibitem[\protect\citeauthoryear{Heil, Vaughan, \& Roberts}{2009}]{2009MNRAS.397.1061H} Heil L.~M., Vaughan S., Roberts T.~P., 2009, MNRAS, 397, 1061 
\bibitem[\protect\citeauthoryear{Ingram \& Done}{2011}]{2011MNRAS.415.2323I} Ingram A., Done C., 2011, MNRAS, 415, 2323 
\bibitem[\protect\citeauthoryear{Jiang, Stone, \& Davis}{2014}]{2014ApJ...796..106J} Jiang Y.-F., Stone J.~M., Davis S.~W., 2014, ApJ, 796, 106
\bibitem[\protect\citeauthoryear{Karachentsev et al.}{2004}]{2004AJ....127.2031K} Karachentsev I.~D., Karachentseva V.~E., Huchtmeier W.~K., Makarov D.~I., 2004, AJ, 127, 2031 
\bibitem[\protect\citeauthoryear{King}{2009}]{2009MNRAS.393L..41K} King A.~R., 2009, MNRAS, 393, L41 
\bibitem[\protect\citeauthoryear{King et al.}{2001}]{2001ApJ...552L.109K} King A.~R., Davies M.~B., Ward M.~J., Fabbiano G., Elvis M., 2001, ApJ, 552, L109
\bibitem[\protect\citeauthoryear{King \& Lasota}{2014}]{2014MNRAS.444L..30K} King A., Lasota J.-P., 2014, MNRAS, 444, L30 
\bibitem[\protect\citeauthoryear{King \& Muldrew}{2016}]{2016MNRAS.455.1211K} King A., Muldrew S.~I., 2016, MNRAS, 455, 1211 
\bibitem[\protect\citeauthoryear{King, Taam, \& Begelman}{2000}]{2000ApJ...530L..25K} King A.~R., Taam R.~E., Begelman M.~C., 2000, ApJ, 530, L25
\bibitem[\protect\citeauthoryear{Lasota, King, \& Dubus}{2015}]{2015ApJ...801L...4L} Lasota J.-P., King A.~R., Dubus G., 2015, ApJ, 801, L4 
\bibitem[\protect\citeauthoryear{Lyubarskii}{1997}]{1997MNRAS.292..679L} Lyubarskii Y.~E., 1997, MNRAS, 292, 679 
\bibitem[\protect\citeauthoryear{Middleton et al.}{2015}]{2015MNRAS.454.3134M} Middleton M.~J., Walton D.~J., Fabian A., Roberts T.~P., Heil L., Pinto C., Anderson G., Sutton A., 2015, MNRAS, 454, 3134 
\bibitem[\protect\citeauthoryear{Middleton et al.}{2015}]{2015MNRAS.447.3243M} Middleton M.~J., Heil L., Pintore F., Walton D.~J., Roberts T.~P., 2015, MNRAS, 447, 3243
\bibitem[\protect\citeauthoryear{Middleton et al.}{2014}]{2014MNRAS.438L..51M} Middleton M.~J., Walton D.~J., Roberts T.~P., Heil L., 2014, MNRAS, 438, L51 
\bibitem[\protect\citeauthoryear{Middleton et al.}{2013}]{2013Natur.493..187M} Middleton M.~J., et al., 2013, Natur, 493, 187
\bibitem[\protect\citeauthoryear{Middleton et al.}{2012}]{2012MNRAS.420.2969M} Middleton M.~J., Sutton A.~D., Roberts T.~P., Jackson F.~E., Done C., 2012, MNRAS, 420, 2969
\bibitem[\protect\citeauthoryear{Middleton, Sutton, \& Roberts}{2011}]{2011MNRAS.417..464M} Middleton M.~J., Sutton A.~D., Roberts T.~P., 2011, MNRAS, 417, 464 
\bibitem[\protect\citeauthoryear{Middleton et al.}{2011}]{2011MNRAS.411..644M} Middleton M.~J., Roberts T.~P., Done C., Jackson F.~E., 2011, MNRAS, 411, 644
\bibitem[\protect\citeauthoryear{Motch et al.}{2014}]{2014Natur.514..198M} Motch C., Pakull M.~W., Soria R., Gris{\'e} F., Pietrzy{\'n}ski G., 2014, Natur, 514, 198 
\bibitem[\protect\citeauthoryear{Ohsuga \& Mineshige}{2011}]{2011ApJ...736....2O} Ohsuga K., Mineshige S., 2011, ApJ, 736, 2 
\bibitem[\protect\citeauthoryear{Pakull \& Mirioni}{2003}]{2003RMxAC..15..197P} Pakull M.~W., Mirioni L., 2003, RMxAC, 15, 197
\bibitem[\protect\citeauthoryear{Pinto, Middleton, \& Fabian}{2016}]{2016Natur.533...64P} Pinto C., Middleton M.~J., Fabian A.~C., 2016, Natur, 533, 64 
\bibitem[\protect\citeauthoryear{Pooley \& Rappaport}{2005}]{2005ApJ...634L..85P} Pooley D., Rappaport S., 2005, ApJ, 634, L85
\bibitem[\protect\citeauthoryear{Poutanen et al.}{2007}]{2007MNRAS.377.1187P} Poutanen J., Lipunova G., Fabrika S., Butkevich A.~G., Abolmasov P., 2007, MNRAS, 377, 1187 
\bibitem[\protect\citeauthoryear{Ritter \& Kolb}{2003}]{2003A&A...404..301R} Ritter H., Kolb U., 2003, A\&A, 404, 301 
\bibitem[\protect\citeauthoryear{Rosen et al.}{2015}]{2015arXiv150407051R} Rosen S.~R., et al., 2015, arXiv, arXiv:1504.07051 
\bibitem[\protect\citeauthoryear{Servillat et al.}{2011}]{2011ApJ...743....6S} Servillat M., Farrell S.~A., Lin D., Godet O., Barret D., Webb N.~A., 2011, ApJ, 743, 6 
\bibitem[\protect\citeauthoryear{Shakura \& Sunyaev}{1973}]{1973A&A....24..337S} Shakura N.~I., Sunyaev R.~A., 1973, A\&A, 24, 337 
\bibitem[\protect\citeauthoryear{Soria et al.}{2012}]{2012ApJ...750..152S} Soria R., Kuntz K.~D., Winkler P.~F., Blair W.~P., Long K.~S., Plucinsky P.~P., Whitmore B.~C., 2012, ApJ, 750, 152 
\bibitem[\protect\citeauthoryear{Stobbart, Roberts, \& Warwick}{2004}]{2004MNRAS.351.1063S} Stobbart A.-M., Roberts T.~P., Warwick R.~S., 2004, MNRAS, 351, 1063 
\bibitem[\protect\citeauthoryear{Sutton, Roberts, \& Middleton}{2013}]{2013MNRAS.435.1758S} Sutton A.~D., Roberts T.~P., Middleton M.~J., 2013, MNRAS, 435, 1758 
\bibitem[\protect\citeauthoryear{Vaughan et al.}{2003}]{2003MNRAS.345.1271V} Vaughan S., Edelson R., Warwick R.~S., Uttley P., 2003, MNRAS, 345, 1271 

\end{thebibliography}
\end{document}